\documentclass[%
 reprint,
 amsmath,amssymb,
 aps,
]{revtex4-1}

\usepackage{graphicx}% Include figure files
\usepackage{dcolumn}% Align table columns on decimal point
\usepackage{bm}% bold math
\begin{document}
%\linenumbers

\preprint{APS/123-QED}

\title{Escape dynamics through a continuously growing leak}


\author{Tam\'as Kov\'acs}
\email[]{tkovacs@general.elte.hu}
%\thanks{Fulbright Fellow}
%\altaffiliation{Department of Astrophysical Sciences, Princeton University, 08544 Princeton, NJ}
\affiliation{Institute of Theoretical Physics, E\"otv\"os University,
P\'azm\'any P. s. 1A, H-1117 Budapest, Hungary and\\
Konkoly Observatory,
Research Centre for Astronomy and Earth Sciences,
Hungarian Academy of Sciences,
H-1121, Budapest, Konkoly Thege Mikl\'os  \'ut 15-17, Hungary
}

\author{J\'ozsef Vany\'o}
%\affiliation{Eszterh\'azy K\'aroly University of Applied Sciences, Faculty of Natural Sciences, H-3300, Eger, Hungary}
\affiliation{Eszterh\'azy K\'aroly University, % of Applied Sciences,
Faculty of Natural Sciences, H-3300, Eger, Hungary and\\
Konkoly Observatory,
Research Centre for Astronomy and Earth Sciences,
Hungarian Academy of Sciences,
H-1121, Budapest, Konkoly Thege Mikl\'os  \'ut 15-17, Hungary
}

\date{\today}

\begin{abstract}
We formulate a model that describes the escape dynamics in a
leaky chaotic system in which the size of the leak depends on the
number of the in-falling particles.
The basic motivation of this work is the astrophysical process which describes the planetary accretion.
In order to study the dynamics generally, the standard map is investigated in two cases
when the dynamics is fully hyperbolic and in the presence of KAM islands.
In addition to the numerical calculations, an analytic solution to the temporal
behavior of the model is also derived.
We show that in the early phase of the leak expansion, as long as
there are enough particles in the system, the number of survivors deviates from
the well-known exponential decay.
Furthermore, the analytic solution returns the classical result in
the limiting case when the number of particles does not affect the leak size.\\
PACS NUMBERS, AND KEYWORDS
\end{abstract}


\maketitle

%\section{}
%\subsection{}
%\subsubsection{}

\section{Introduction \label{sec:Intro}}

Simple nonlinear dynamical systems in which trajectories may escape
through an artificial leak
\footnote{Artificial means in this context that if the leak is not present, escape cannot occur at that part of the phase space.}
placed in the phase space play an important role in recent studies.
Various fields of physics deal with either the escape dynamics of the particles or the decay rate of other physical quantities such as sound intensity, light rays, or fractal eigenstates \cite{Sch2002,Jun1993,Ern2014,Nag2005,Alt2009,Has2013,Por2008}.
It has been pointed out that the escape dynamics strongly depends on
the leak size, position, and orientation
\cite{Lai1999,Zyc1999,Afr2010,Bun2007,Det2012,Det2013,Det2011} as
well as on other pre-defined properties of the leak, for instance, the reflection coefficient \cite{Alt2013}.
Probably the most interesting question is how the escape dynamics changes if the size
of the leak varies.
Altmann et al. \cite{Alt2013R} presented numerical results about the relation between the escape rate and the leak size.
In their study, however, the measure of the leak was adjusted manually in each case.
Recently, Livorati et al. \cite{Liv2014} studied the escape in case of periodically
driven holes.
The main results of their work show parameter
(amplitude, initial phases, and period of the oscillations)
dependent fluctuations superimposed to the classical exponential decay.

Although mathematicians are interested mostly in the limiting case of vanishing small leaks \cite{Hay2005,Kel2009,Szasz2012}, in this work we present the decay dynamics through a continuously growing leak, where the size of the leak depends on a given physical property of the escaping particles.
The motivation of this study comes from the application of leaky chaotic systems
\cite{Ass2014,Kov2015,Mor2015} and crash tests \cite{Zot2015,Zot2016a}
in dynamical astronomy discussed in details below.

The model of the growing leak introduced here results in a survival probability of non-escaped trajectories that is different from the well-known classical exponential decay \cite{Lai2011,Tel2006}.
Moreover, we found a simple analytical solution describing the escape dynamics until the leak's expansion stops.
A comprehensive numerical investigation is also performed to confirm our analytic results.

The paper is organized as follows.
After the Introduction, in Section \ref{sec:model}, the motivation as an astrophysical
application is described.
Then, we give a detailed description of the model of a growing leak and its simple numerical implementation to the standard map.
The mathematical background
%(the differential equations, the analytic solution,
%and the calculation method of distributions)
is presented in Sections \ref{sec:averaged_behaviour}.
Section \ref{sec:num_res} is devoted to numerical calculations
in order to compare analytic results and simulations.
Finally, we discuss our results and draw some conclusions in Section \ref{sec:sum_disc}.

\section{Model \label{sec:model}}

\subsection{Motivation \label{sec:rtbp}}

The motivation of the present study \cite{Gol1973} is the so-called planetary accretion process which is one of the two competing planet formation scenarios in these days \cite{Mat2007}.
%\footnote{The other one is the disk instability model. For more information
%see e.g.}.
In this process the forming planetary embryo accretes particles from its vicinity until this region -- the \textit{feeding zone}
\footnote{The planetary feeding zone is basically the basin of attraction of a given
leak where the leak in phase space can be considered as the forming planetesimal.}
-- becomes empty.
%The mass the planet can collect in this way is called the isolation mass.
The increase of the planet depends on the mass of the particles hitting its surface.
Obviously the smaller the embryo at the beginning of this process,
the more significant the growth by the accretion.
As a very simple model of this process one might consider the gravitational planar circular restricted three body problem (RTBP).
In RTBP two point masses (star and planet) orbiting their barycenter on a circle and a
third mass-less body (test particle) moves in their gravitational potential in
the same plane.
Although the planet (and also the star) is considered as a point mass, one can define
the Hill radius ($r_{\mathrm{H}}$) in which its gravitational influence is dominant.
The particles entering the Hill radius with an appropriate velocity, i.e. slower than the escape velocity from this domain, can be removed from the dynamics and marked as escaped.
{In addition, $r_{\mathrm{H}}$ grows with the mass of the forming planet, see Eq.~(\ref{eq:rH}).}
Therefore, the growth of the planetary embryo can be considered as a growing
leak in the phase space.
Thus, from dynamical point of view, the accretion stage of the planet formation can be
described via leaky chaotic systems.
%\textbf{The derivation of the exact
%formula of the leak size depending on mass in RTBP is postponed to the Appendix
%\ref{appendix:size_mass}. The mass of the leak depends on the mass of the system. We will give an estimate on the size of the leak in Appendix \ref{appendix:size_mass}.
We give an estimate how the leak size depends on the mass in RTBP, see Appendix \ref{appendix:size_mass}.

\begin{figure}[b]
\includegraphics[width=0.475\textwidth]{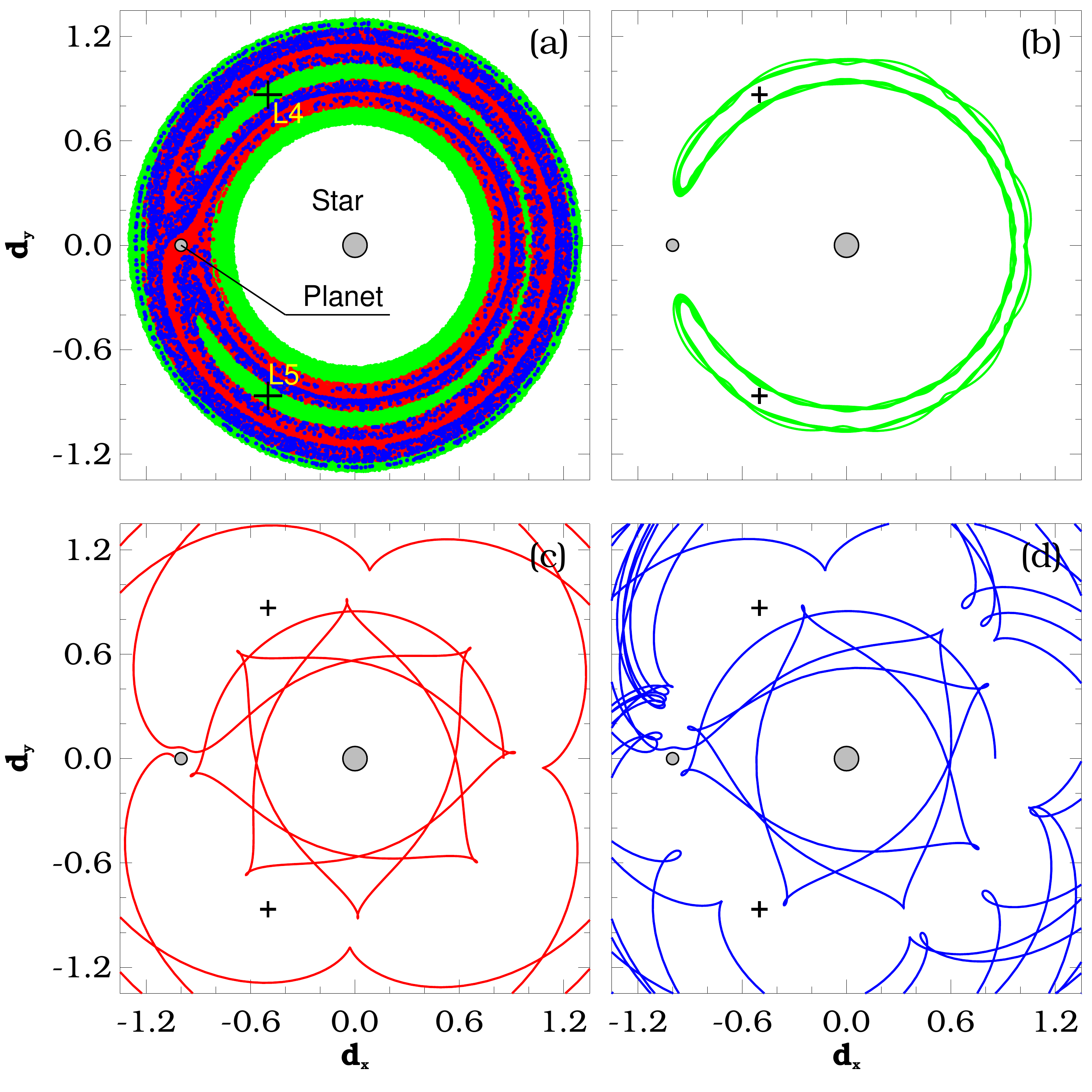}
\caption{\label{fig:rtbp} (color online)
(a) An example of a growing leak in dynamical astronomy.
%The size of a forming planet can be described as a 4D ($x,\;y,\;v_x,\;v_y$)
%leak in the phase space of the RTBP.
The plot shows initial conditions from the annulus around the planet's orbit.
The end-states of the particles are color coded as described in the text.
%Light gray (Green): trajectories remaining bounded during the whole integration;
%Gray (red): particles hitting the planet's surface;
%Dark gray (blue): trajectories scattered out due to the gravitational perturbation of the planet.
Particles have been started on circular Keplerian orbit.
The size of the planet and star are enlarged for better visualization.
The triangular Lagrangian points ($L_4\text{ and }L_5$) are also marked.
(b)-(d) Examples for individual orbits corresponding to certain initial conditions
in panel (a).
%Colors indicate the fate of the particles.
Note that the end point of the light gray (blue) trajectory is outside the plotted region.
}
\end{figure}

To illustrate the leaky RTBP,
we plot the evolution of a large number of non-interacting test particles initially placed
around the planet's orbit (see Figure \ref{fig:rtbp}).
Different colors denote different end-states of particles.
Trajectories starting from light gray (green online) points remain the part of the system during the whole
integration (1000 orbits of the planet).
Gray (red online) points represent test particles whose destination is the planet, more precisely,
the half of the Hill radius with proper velocity
\footnote{Since we consider the planet as a point mass, crash of the particles and the
planetary embryo is difficult to calculate numerically.
Consequently, half of the Hill radius is chosen as a region wherein
the particles are thought to be accreted by the forming planet.
This is, obviously, more rigorous criterion than one Hill radius.}.
Dark gray (blue online) points indicate trajectories scattered out from the system by the planet.

Although the effect of the planet's mass and size evolution in the RTBP is
dominant only in very early stages of the planet formation, the idea of a growing leak,
particularly when the size of the leak depends on a physical property of the leaving
particles, might shed light on a new kind of escape dynamics generally in leaky chaotic systems.

\subsection{Growing leak model\label{sec:leak}}

%Let us consider a discrete-time dynamical system which represents
%the trajectory of the particles in the phase space of a real physical system.

{
%Let us consider a discrete-time dynamical system
%which represents the phase space of a real physical system.
%Let us consider a discrete-time dynamical system.
%The evolution of its state represents the trajectory of a particle moving
%in real physical system.
%A great number of particles are moving in the system at the same time.
%There are a great number of particles in the system at the same time.
%In the phase space there is a domain -- the so called leak -- through
%which the particles can escape.
The discrete dynamical system we are to consider here consists a large number of
particles and a leak, where under certain conditions, the particles can
escape from the system.
}
The particles are point masses with the same mass $m$,
their initial number is $N_0$,
while after $i$ iterations we denote the number of surviving particles by $N_i$.
The leak also has an initial and an instantaneous mass, $M_0$ and $M_i$, respectively.
When a particle falls into the leak, its mass is added to that of the leak, thus
\begin{eqnarray}
M_i = M_0 + (N_0-N_i) \cdot m = \mathcal{M} - N_i \cdot m,
\label{eq:leakmass}
\end{eqnarray}
where $\mathcal{M} = M_0 + N_0 \cdot m$ is the total mass of the system.

{According to the RTBP (Appendix~\ref{appendix:size_mass}) a reasonable choice is }that the volume of the leak depends on its mass $M_i$ in the form of
\begin{eqnarray}
S_{\mathrm{leak}}(M_i) = C_{S} \cdot M_i^\gamma,
\label{eq:T_leak}
\end{eqnarray}
{where $\gamma$ is a positive constant. The coefficient $C_S$ can be written as $C_S=C_P\cdot S_{\mathrm{total}}$.
Here $C_P>0$ denotes a normalization constant while $S_{\mathrm{total}}$ is the
{volume} of the ergodic part of the phase space.}
The factor $C_P$ allows us to control the final size of the leak,
(a leak of moderate size avoids excessive restructuring of the phase space).
Let $p$ be the escape probability that a particle leaves the system
(through the leak) in the next iteration.
We {suppose} that the escape probability is proportional to the actual size of the leak compared to the whole phase space, that is,
$p=S_{\mathrm{leak}}/S_{\mathrm{total}}.$
That is, the escape probability (see Eq.~(\ref{eq:T_leak})) is given by
\begin{eqnarray}
p(M_i) = C_p \cdot M_i^\gamma. \label{eq:escape_prob_M}
\end{eqnarray}
%where $C_p$ and $\gamma$ are positive constants.
Generally, the escape probability is changing as the mass (and size) of the leak is increasing.

At this point, it is useful to introduce some new constants and variables:
\begin{eqnarray}
\mathcal{N} = \frac{\mathcal{M}}{m},\;\;\;\;
\kappa_\infty = C_p \mathcal{M}^\gamma,
\nonumber
\label{eq:parameters}
\end{eqnarray}
\begin{eqnarray}
%x(t) = \frac{M(t)}{\mathcal{M}},\;\; x_0 =
%\frac{M_0}{\mathcal{M}},\;\; y(t) = \frac{N(t) \cdot m}{\mathcal{M}}
%= \frac{N(t)}{\mathcal{N}},\;\; y_0 = \frac{N_0}{\mathcal{N}}.
x_i = \frac{M_i}{\mathcal{M}},\;\;\;\;\;
y_i = \frac{N_i}{\mathcal{N}} =
N_i \frac{m}{\mathcal{M}},
\label{eq:norm_functs} \nonumber
\end{eqnarray}
where $\mathcal{N}$ is the number of particles corresponding to the total mass $\mathcal{M}$,
$\kappa_{\infty}$ {is the asymptotic escape rate} when all the mass of the system is in the leak,
{$x$ is the ratio of the mass of the leak and the total mass (mass ratio),
$y$ is the ratio of the number of the particles which are outside the leak to the total number of the particles $\mathcal{N}.$
%and, finally, $x_0=x(0)$ and $y_0=y(0).$% are the initial values of $x(t)$ and $y(t),$ respectively.
It is obvious that
\begin{eqnarray}
x_i + y_i = 1 \nonumber %\;\;\; \mathrm{and}\;\;\; x_0 + y_0 = 1. \nonumber
\end{eqnarray}
for all time instant.
}
We will use these dimensionless quantities through the rest of the paper.
%}

{
The assumption of a small leak in our model corresponds to the pure exponential survival probability, i.e. when the system shows strong chaotic properties.
That is, if a static leak with size equal to the final size of the evolving leak
(set by $C_{\mathrm{P}}$) produces exponential decay,
we consider that this measure of the leak is small enough to our purposes and fits to the zero order approximation
%$p_n=S_{\mathrm{leak},n}/S_{\mathrm{total}},$
$p=S_{\mathrm{leak}}/S_{\mathrm{total}},$
widely used in the literature, see for example \cite{Det2009}.
In addition, the exponential decay can also be observed in weakly chaotic systems for short times until the hyperbolic dynamics dominate.
}

{
Furthermore, in case of weak chaos the growing leak in the model presented should avoid the quasiperiodic domain in the phase space.
On the other hand, if the leak intersects the KAM tori during its growth, the survival probability will decay with lower different rate.
In other words, since the regular domain behaves as a forbidden region for trajectories originating outside, the leak biting into it will have an unreachable part for those trajectories resulting in a different escape probability.
However, this is no longer true when the leak originally contains islands or more precisely when the ratio of the regular islands inside and outside the leak remains constant.
}

\subsection{Simplified numerical experiment\label{sec:stmap}}

In order to analyze the escape dynamics through a continuously growing leak
defined by Eq.~(\ref{eq:escape_prob_M}), we introduce a simple test system.
Our numerical experiments are based on the standard map ($\text{mod}\; 2\pi$) which
describes the Poincar\'e map of the kicked rotator.

\begin{figure}[b]
\includegraphics[width=0.475\textwidth]{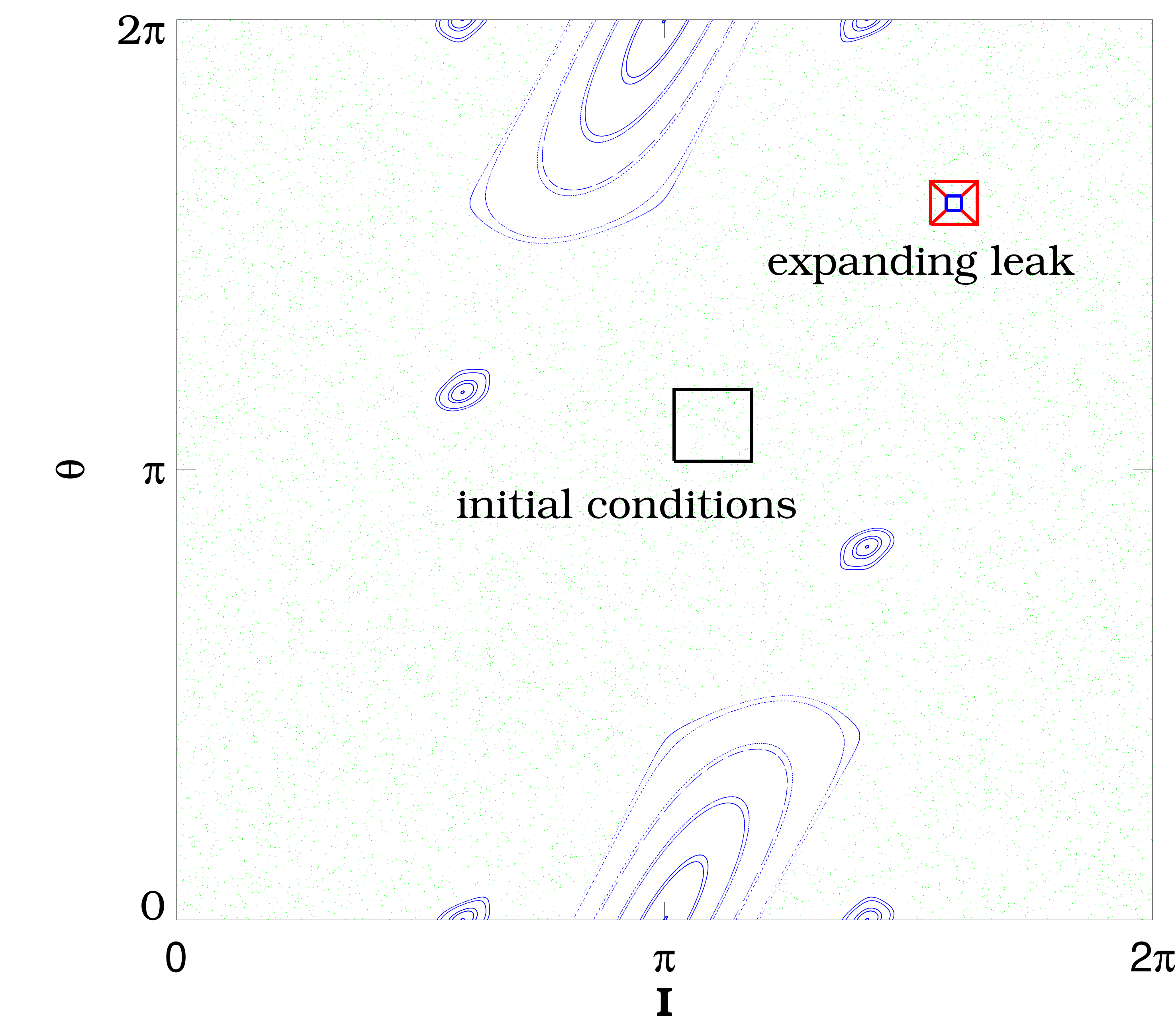}
\caption{\label{fig:SM}
Visualization of the numerical setup.
%The phase space portrait of the SM for $K$=2.7 is shown, the box in the
%middle (black) contains the initial conditions while the expanding leak is
%placed at $(\Theta,I)=$(5,5) (red on-line).
The invariant curves (blue online), plotted for completeness, are related to
different initial conditions than those show by dots representing the chaotic trajectories.}
\end{figure}

This choice makes it possible to check the leak's expansion in both co-ordinate
and velocity directions, respectively.
The standard map (SM) reads as follows
\begin{equation}
\begin{split}
I_{i+1}&=I_{i}+K\sin \Theta_{i},\\
\Theta_{i+1}&=\Theta_{i}+I_{i+1}.
\end{split}
\label{eq:SM}
\end{equation}
In Eq.~(\ref{eq:SM}) $K$ denotes the strength of the perturbation
and allows to study either \textit{fully hyperbolic dynamics}
($K$=5.19) or \textit{mixed phase space} structure, e.g. $K=2.7$.
{
An other reason we consider the SM is that it allows us to mimic the conservative dynamics in the RTBP where regular islands are also embedded in the chaotic sea producing the well-known structure of the phase space similar to that in Fig.~\ref{fig:SM}.
}

For simplicity, we presume that the leak grows equally in $I$ and $\Theta$ directions,
i.e. it conserves its original shape.
In order to avoid the early irregular effects in escape rate due to the location and
density of the initial conditions, a threshold time is obtained before the leak is opened.
Thus, we have a uniform distribution of the trajectories in the ergodic region of the phase space.
The threshold time is set to be $i=250$ in all simulations.

Figure \ref{fig:SM} shows the phase space portrait of the SM for $K$=2.7.
%As an example,
We place a square-shaped leak centered at point
$(I, \Theta)=$(5,5) with initial size $(\Delta I,\Delta \Theta)=(0.01,0.01)$
($S_{leak}^{(0)}=10^{-4}$)
%mass $x_0=1$
\footnote{
Note that the mass and the size of the leak are identical parameters of the problem.
The instantaneous size can be obtained from the current mass and vice versa.
%That means the leak initially covers the
%$(\Delta I,\Delta \Theta)=(0.01,0.01)$
%area in the phase space.
}
and store the number of escaped trajectories at every iteration step.
The semi-diagonals indicate the expansion until the leak reaches its final size
$(\Delta I,\Delta \Theta)\approx(0.316,0.316)$ ($S_{leak}^{(\infty)}=0.1$).
Initial conditions are placed uniformly in the black square
($3.2\leq \Theta\leq 3.7$, $3.2\leq I\leq 3.7$)
far from KAM islands as well as the final leak.
\begin{figure}[t]
\includegraphics[width=0.475\textwidth]{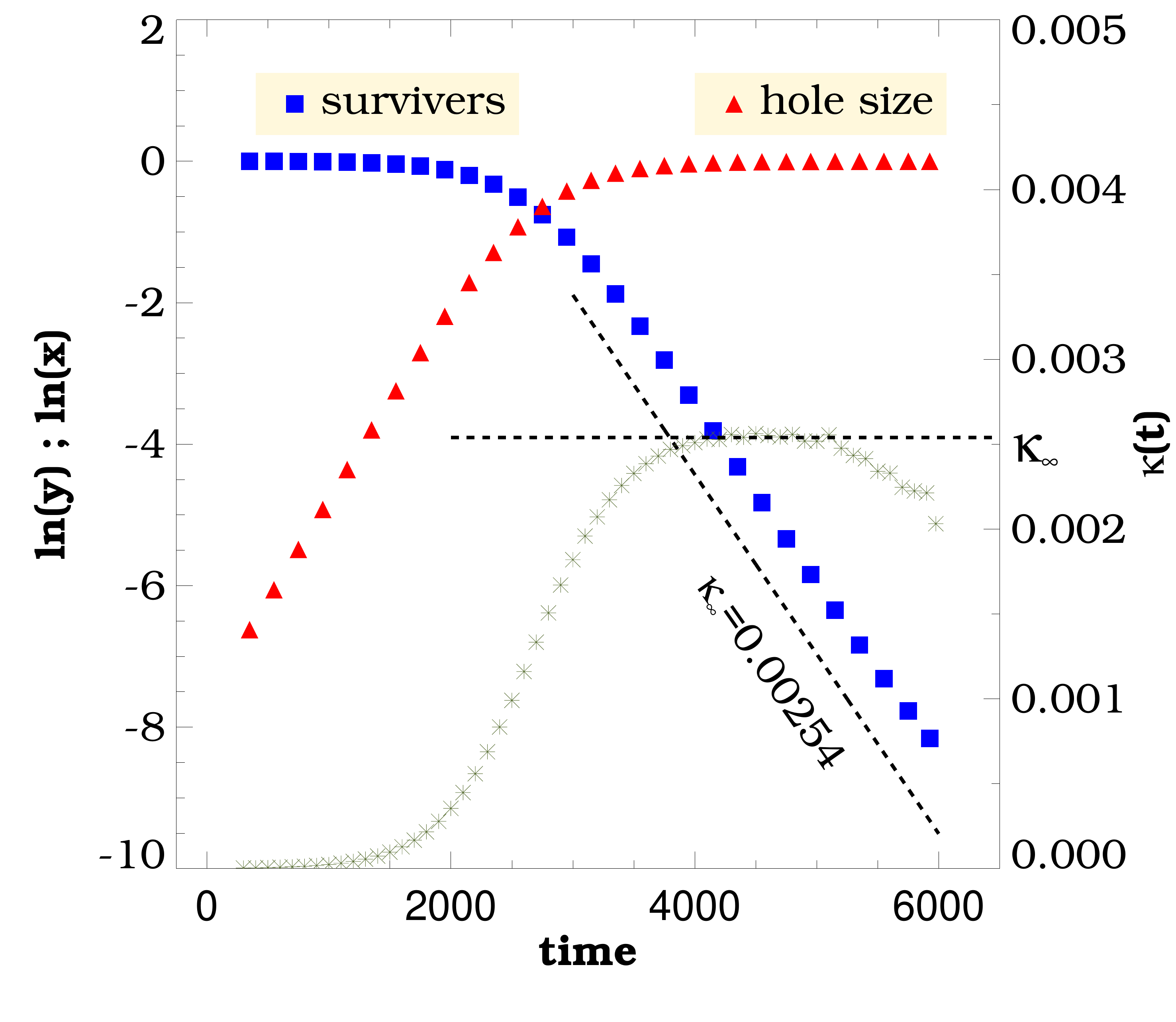}
\caption{
\label{fig:kappa_t}
Escape dynamics in SM.
{
Parameters of the simulation:
$K=2.7$, $\gamma=1$, $m=1$, $C_p = 10^{-7}/(2 \pi)^{2}$, $N_0 = 10^6$, and $M_0 = 1000$.
%Initial conditions are $y_0\approx1,\;x_0\approx 0.$
The leak reaches its final mass at $t\approx 6000.$}
%In this simulation the particles are identical, each has a mass $m=1$
%and $\gamma=1/2$ in Eq. (\ref{eq:escape_prob_M}).
%, the link between the leak's size and mass,
%reads $S_{\mathrm{leak},n}\sim M_n^{1/2}.$
For more details see the text.
}
\end{figure}

The result of a test run is displayed in Figure \ref{fig:kappa_t}.
It is clearly visible that the well-known exponential decay of the non-escaped trajectories starts after $\mathtt{\sim}$3000 iterations (blue squares).
Red triangles denote the instantaneous leak size, $S_{\mathrm{leak}}$, which is growing rapidly until it reaches its final ($\mathtt{\sim}$90\%) size.
One can also observe that the exponential decay starts roughly when the expansion of the leak ceases.
We can, thus, presume that the exponential behavior is a consequence of the stationary leak size with escape rate $\kappa_{\infty}$.
{
The semi-logarithmic plot of the non-escaped trajectories allows one to find the asymptotic escape rate, $\kappa_{\infty}$ as $t\to\infty$ (for strong chaotic regime).
%In this study we denote the escape rate, corresponding to the stationary leak size,
This simulation yields $\kappa_{\infty}=0.00254$.
%and final leak size $S_{\mathrm{leak}}\approx 0.1$
%(i.e. linear size $\Delta I=0.314$, $\Delta \Theta=0.314$).}
}

Furthermore, the numerical investigation confirms the naive idea that until the leak's expansion is present, the {instantaneous} escape rate, $\kappa(t),$ and also the escape probability is changing in time according to
$\mathrm{d}(\ln y_n)/\mathrm{d}t=-\kappa(t)$.
However, when the growth slows down significantly $\kappa(t)$ reaches the {asymptotic} escape rate $\kappa_{\infty}$ (green asterisks),
see Figure \ref{fig:kappa_t}.
This behavior can be explained as follows.
At the beginning of the simulation ($t<2000$) a very large number
of escaping trajectories feed the small leak in one iteration step
and, therefore, its mass (size) growth is accelerating.
Beyond a certain limit the mass (or equivalently the number) of escaping particles
in one iteration compared to the mass of the leak becomes small,
i.e. escape is present with moderate increase of the leak size.
In this case ($2000\leq t\leq 5000$), however, there are enough
particles in the system to observe the exponential decay.

The reason for the larger dispersion in $\kappa(t)$ and its deviation
from $\kappa_{\infty}$ beyond $t\approx 5000$ is twofold.
On the one hand, the number of non-escaped trajectories, after 5000 iterations,
becomes so small ($\sim$100) that the statistic is unreliable.
On the other hand, Figure \ref{fig:kappa_t} shows the
simulation for $K$=2.7, in which case KAM tori are responsible for
stickiness and consequently a power-law decay of trajectories for
longer escape times (not shown).
In other words, $\kappa(t)$ would follow the horizontal dashed line in
case of the fully hyperbolic dynamics, for instance, $K\ge$~5.19, with an arbitrarily large $N_0.$

\section{Results \label{sec:results}}

\subsection{Analytic solution \label{sec:averaged_behaviour}}

After having some impression about the escape dynamics from numerical simulations,
in this section, we show that a continuous approximation of the temporal
behavior of the model can be described by analytic formulae.
{
We consider the particle number $N_i$
%,the mass of the leak $M_i$ and all other functions corresponding to them
and all the other related discrete functions $M_i$, $x_i$, and $y_i$
as being continuous functions $N(t)$, $M(t)$, $x(t)$, and $y(t)$.
}
Practically, we can do that because the particle number and the typical timescale
(number of iterations) of the process is also much higher than unity ($N_0\gg1$).

The time derivative of $N(t)$ is approximately the negative of
the average number of escaping particles $\Delta N$ during one
iteration which is $p \cdot N$, so we can write
\begin{eqnarray}
\frac{\mathrm{d}N}{\mathrm{d}t}\approx \Delta N =
- p \cdot N = - C_p \cdot M^\gamma \cdot N
\label{eq:N derivative}
\end{eqnarray}
where we used Eq.~(\ref{eq:escape_prob_M}).
%Similarly, the time derivative of $M(t)$ is
As $\Delta M = - \Delta N \cdot m$, the time derivative of $M(t)$ is
\begin{eqnarray}
\frac{\mathrm{d}M}{\mathrm{d}t}\approx
%\Delta M = - \Delta N \cdot m =
C_p \cdot M^\gamma \cdot N \cdot m.
\label{eq:M derivative}
\end{eqnarray}
Combining Eq.~(\ref{eq:M derivative}), $M(t) = x(t) \cdot \mathcal{M}$, and
$N(t) = y(t) \cdot \mathcal{N}=(1-x(t)) \cdot \mathcal{M}/m$,
we get a first-order separable ordinary differential equation for $x(t)$:
\begin{eqnarray}
\frac{\mathrm{d}x}{\mathrm{d}t}
=
\kappa_{\infty} \cdot x^\gamma \cdot \left( 1 - x \right),
\label{eq:ode for x}
\end{eqnarray}
%where dot denotes the time derivative.
% and $x(0)=x_0.$
%The solution of Eq.~(\ref{eq:ode for x}) can be found in Appendix~\ref{appendix:solution_of_eq}.
Derivation of the solution can be found in Appendix~\ref{appendix:solution_of_eq}.
{
Equation (\ref{eq:ode for x}) is a continuous approximation of the recursive difference equation
\begin{eqnarray}
x_{i+1}=x_i + \Delta x_i
%\Delta x_i = \kappa_\infty x_i^\gamma \left( 1 - x_i \right)
%\Leftrightarrow
%x_{i+1}=x_i + \kappa_\infty x_i^\gamma \left( 1 - x_i \right)
\label{eq_discrete}
\end{eqnarray}
%where $x_{i+1}=x_i + \Delta x_i$
where $\Delta x_i = \kappa_\infty \cdot x_i^\gamma \cdot \left( 1 - x_i \right)$,
which gives the exact description of the discrete-time problem.
}

The implicit solution of (\ref{eq:ode for x}) can be given by
\begin{eqnarray}
t(x)=
\frac{x^{1-\gamma}}{\kappa_{\infty} \cdot (1-\gamma)}\;
{}_2F_1\left(1-\gamma,1;2-\gamma;x\right) - \tau
\label{eq:sol}
\end{eqnarray}
where the constant of integration $\tau$ follows from the initial value $x_0$ as
\begin{eqnarray}
\tau = \frac{x_0^{1-\gamma}}{\kappa_{\infty} \cdot (1-\gamma)}\;
{}_2F_1\left(1-\gamma,1;2-\gamma;x_0\right).
\label{eq:int_const}
\end{eqnarray}
\begin{figure}
\includegraphics[width=0.95\linewidth, angle=0]{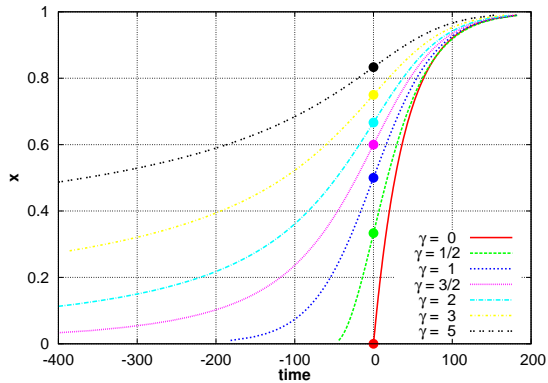}
\caption{ The mass growth of the leak $x(t)$ for different $\gamma$s.
For better visibility, the constants of integration ($\tau$) are
chosen with $x_0=x_{\mathrm{PoI}}$ taken at 0 see Eq.~(\ref{eq:int_const}).
Parameter $\kappa_{\infty}$ is taken equal to $1/(2\pi)^{2} \approx 0.0253$.
\label{fig:analytical} }
\end{figure}

The solution of Eq.~(\ref{eq:ode for x}), $x(t),$ has a point of inflection (PoI) for all $\gamma>0.$
The second derivative of $x$ from (\ref{eq:ode for x})
\begin{eqnarray}
\frac{\mathrm{d}^2x}{\mathrm{d}t^2} =
\kappa_{\infty} \cdot x^{\gamma-1} \cdot \frac{\mathrm{d}x}{\mathrm{d}t}
\cdot \left[\gamma - ( 1 + \gamma )\cdot x \right],
\nonumber
\end{eqnarray}
from which the $x$ coordinate of the inflection point
($x_{\mathrm{PoI}}$)
can be obtained %. It is
\begin{eqnarray}
x_{\mathrm{PoI}} =  \frac{\gamma}{1 + \gamma}.
\end{eqnarray}
We further elaborate on the error properties of the above solution in Appendix \ref{appendix:error}.

We can distinguish two parts of the leak-growing process.
The separatrix is the point of inflection of the $x(t)$ function.
Figure \ref{fig:analytical} shows the functions $x(t)$ for different $\gamma$s.
For the sake of comparison the graphs are shifted leftward, thus,
the inflection points are placed exactly above a row at $t=0.$

The mass growth $x(t)$ beyond the point $x_{\mathrm{PoI}}$ (or $t=0$) has
the same characteristic for different $\gamma$s.
The reason is that in the limit $t \rightarrow \infty$, $x \rightarrow 1,$ Eq.~(\ref{eq:ode for x})
can be written as
%$\dot{x} \approx -\kappa_{\infty}x$
$\mathrm{d}x/\mathrm{d}t \approx -\kappa_{\infty}x$
which means that function $x(t)$ approximates 1 exponentially with exponent
$-\kappa_{\infty}$ and the process does not depend on $\gamma.$

This is, however, not the case to the left of the point of inflection.
In the limit of $x\rightarrow 0$, Eq.~(\ref{eq:ode for x}) can be written as
$\mathrm{d}x/\mathrm{d}t \approx \kappa_{\infty}x^\gamma$ which means that the solution
$x(t)\approx\left[\kappa_{\infty} (1-\gamma) (t +\tau)\right]^{\frac{1}{1-\gamma}}$
follows a power-law and contains both $\kappa_{\infty}$ and $\gamma.$

Furthermore, in this regime $\gamma$ defines two different behaviors.
Considering the case of $\gamma<1$ we have a point where $x(-\tau)=0.$
That is, the integration constant $\tau$ is suitable to determine a time instant
in the past when the mass of the leak was zero, i.e. when the whole growing process began.
While in the case of $\gamma \ge 1$ the function $x(t)$ approaches zero only
in the limit of $t \to -\infty.$
In summary
\begin{eqnarray}
\lim_{x\rightarrow 0} t = \left\{
 \begin{array}{ll}
  -\tau &\;\;\; \textrm{for $0\le\gamma<1$,}\\
  -\infty &\;\;\; \textrm{for $1\le\gamma$.}
 \end{array}
\right.
\label{eq_limit_of_t}
\end{eqnarray}
%{\bf
%?????????? In the case of $\gamma<1$ the picture is a little bit more complicated as it
%is showed in Appendix \ref{appendix:error}.
%}

Nevertheless, it is obvious from Eq.~(\ref{eq:ode for x}) that $\kappa_{\infty}$ is inversely proportional to the timescale of the process.
The condition that the timescale have to be much higher than unity is equivalent
to $1/\kappa_{\infty} \gg 1$.
%(in practice $1/\kappa_{\infty} > 100$).
This fact is important to ensure that the continuous time approximation,
Eqs.~(\ref{eq:N derivative}) and (\ref{eq:M derivative}), is valid in our model.

The adopted model of growing leak defines a stochastic process,
whose complete description is possible only by using the probability theory.
The question arises naturally,
how the probability mass function of the particle number
can be calculated after the $i$th iteration if the initial one is known?
The question is important because if the standard deviations are considerable,
then we need the probability mass functions in order to have a complete description.
Otherwise, the averaged behavior, studied previously, describes the process well.
In Appendix \ref{appendix:distr} we derive the probability mass functions, and study
its properties this problem.
%We find that the standard deviations remain small.

We should mention that during the calculation we assumed that $\gamma>0.$
However, it is obvious that solutions of Eq.~(\ref{eq:ode for x}) can also be found
for negative exponents in a similar way.
The discussion of the case $\gamma<0$ is beyond the scope of the present study.

%{\bf
%Our closing remark in this section concerns
%the similar leaky but continuous-time dynamical systems.
%For these kinds of systems
%the Eq. (\ref{eq:ode for x}) and its solution (\ref{eq:sol}) are not only approximations but give the exact description of the process.
%If the time is continuous in the problem then the Eq. (\ref{eq:ode for x})
%is not an approximation but it describes the behaviour exactly.
%}

\subsection{Numerical tests \label{sec:num_res}}

After discussing the analytic description of the survival probability,
we confirm the validity of our calculations by running several numerical simulations.
In order to demonstrate the general phenomenon of escape dynamics,
we use different $\gamma$ values in our calculations.
%They are always specified either in text or in figure captions.

\begin{figure}[b]
\includegraphics[width=0.475\textwidth]{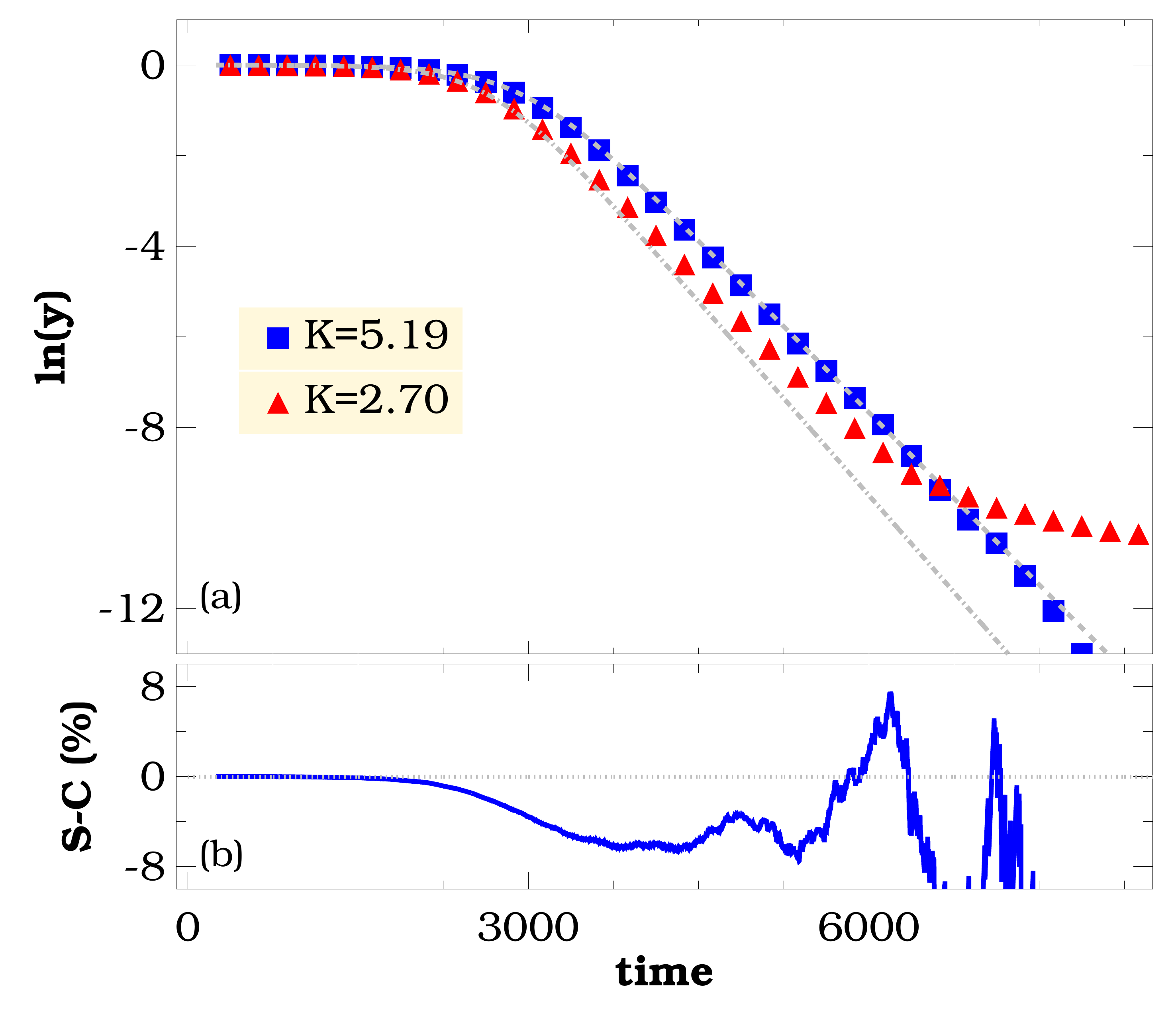}
\caption{\label{fig:gamma1}
{
(a) Survival probabilities for different parameters $K=5.19$ and $2.7$ in SM.
Parameters of the simulation:
$\gamma=1$, $m=1$, $C_p = 10^{-7}/(2 \pi)^{2}$, $N_0 = 10^6$, and $M_0 = 1000$.
%Particles are identical with equal masses, $m_{i}$=1.
The gray dashed ($K=5.19$) and dashed-dotted ($K=2.7$) lines represent the analytic formula
(\ref{eq: sol gamma 1 exp}) %and (\ref{eq: sol gamma 1 exp tau})
with
$\kappa_\infty = 0.1/(2 \pi)^{2} \approx 0.00253$, $x_0 = 10^{-3}$
and
$\kappa_\infty = 0.00285$, $x_0 = 10^{-3}$
respectively.
(b) The difference between the numerical simulation and
the analytic formula
%Eq.~(\ref{eq: sol gamma 1 exp})
for $K=5.19$.
}
}
\end{figure}

\begin{figure}[b]
\includegraphics[width=0.475\textwidth]{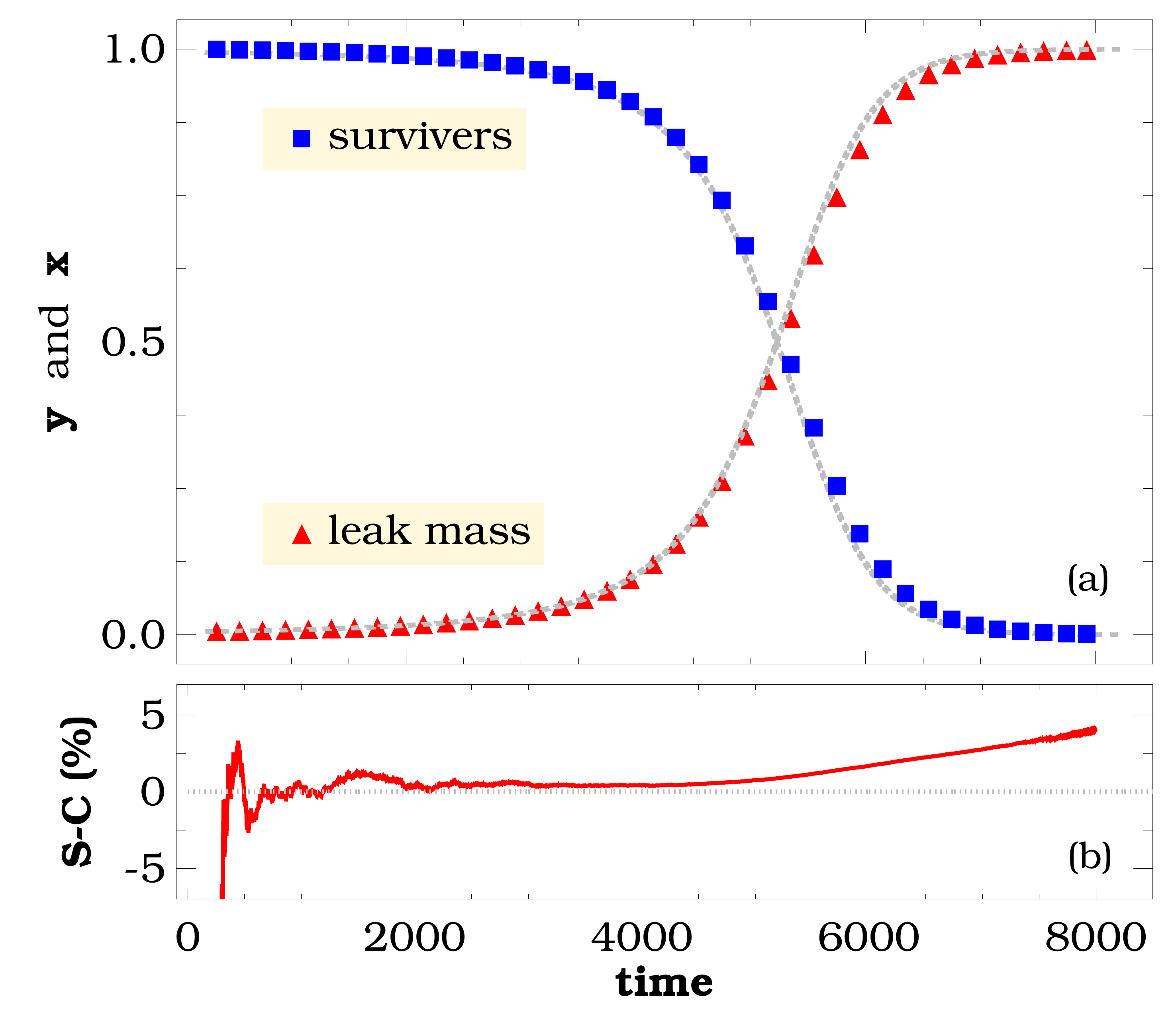}
\caption{\label{fig:gamma43}
(a) The growth of the leak's mass and the decay of particles for $K=2.7$.
The other parameters are
$\gamma=4/3$, $m=1$, $C_p = 0.2845$, $N_0 = 10^6$, and $M_0 = 5631$.
The dashed lines illustrate the analytic solutions
with
$\kappa_\infty = 0.1/(2 \pi)^{2} \approx 0.002533$, $x_0 = 0.0055$
.
(b) $S-C$ curve shows the difference in leak mass.}
%For the reason of the irregular
%behavior for short times, see the text.}
\end{figure}

First, the results of the hyperbolic and mixed dynamics are compared.
In this calculation we show that for different system parameters $K=2.7\text{ and }5.19$
the analytical solution works very well.
Figure \ref{fig:gamma1}(a) shows the ratio of non-escaping trajectories $y(t)$
for the $\gamma=1$ case, i.e. the leak size depends linearly on mass.
One can easily see that the analytical solution (dashed and dotted dashed lines) fit the numerical data fairly accurately, especially for small iteration numbers, $t<2000.$
In order to be able to compare the accuracy of the results quantitatively,
we calculate the relative difference between the simulated data (S) and
the analytic solution (C).
The difference $S-C$ in percentages is plotted in Figure \ref{fig:gamma1}(b).
It shows the same tendency
what we can observe by naked eye in panel (a).
The $S-C$ diagram remains under 4\% level until $t\approx 2500.$
In addition, $S-C$ shows that in the case of $\gamma=1$ the analytic solution is more accurate for fully hyperbolic dynamics ($K=5.19$) than for mixed phase space ($K=2.7$)
for $t>2500$.
The reason of that comes form Eq.~(\ref{eq:sol gamma 1}), since it turns to be purely
exponential for $t\gg 1$, that is
$y(t)\sim exp(-\kappa_{\infty} t)$.
In addition, the decay of $y(t)$ in the latter case starts to deviate from the exponential due to the sticky effect of the KAM tori.

Physically more interesting cases are when $\gamma\neq1$ but rational.
Let us recall our motivation, the planet formation analogy in the planar RTBP.
The size of the leak in the phase space in this particular case is
proportional to $m_p^{4/3},$ see Eq.~(\ref{eq:hole2D}) in the Appendix.
%can be described
%by $S_{\mathrm{leak}}=C_{T}M^{4/3},$ see Eq.~(\ref{eq:hole2D}) in the Appendix.

Figure~\ref{fig:gamma43}(a) shows the number of surviving particles, $x(t),$
and the mass growth of the leak, $y(t),$ for $\gamma=4/3$
(squares and triangles, respectively).
The analytic solution goes together with the numerical simulation also for this value of $\gamma.$
As is well seen in panel (b) the $S-C$ diagram remains under the 5\% level until the leak reaches its final mass, $t\approx8000.$
This is not true, however, at the very beginning of the iteration,
$t<10$ after opening the leak.
In this regime sudden changes in the number of escaping trajectories appear.
Trajectories situated exactly 'above' the leak and its pre-images disappear immediately from the system.
This rapid change in the number of particles is, however, not covered
by the analytic solution and, consequently,
large differences may show up in the first phase of the $S-C$ diagram.

\begin{figure}[b]
\includegraphics[width=0.475\textwidth]{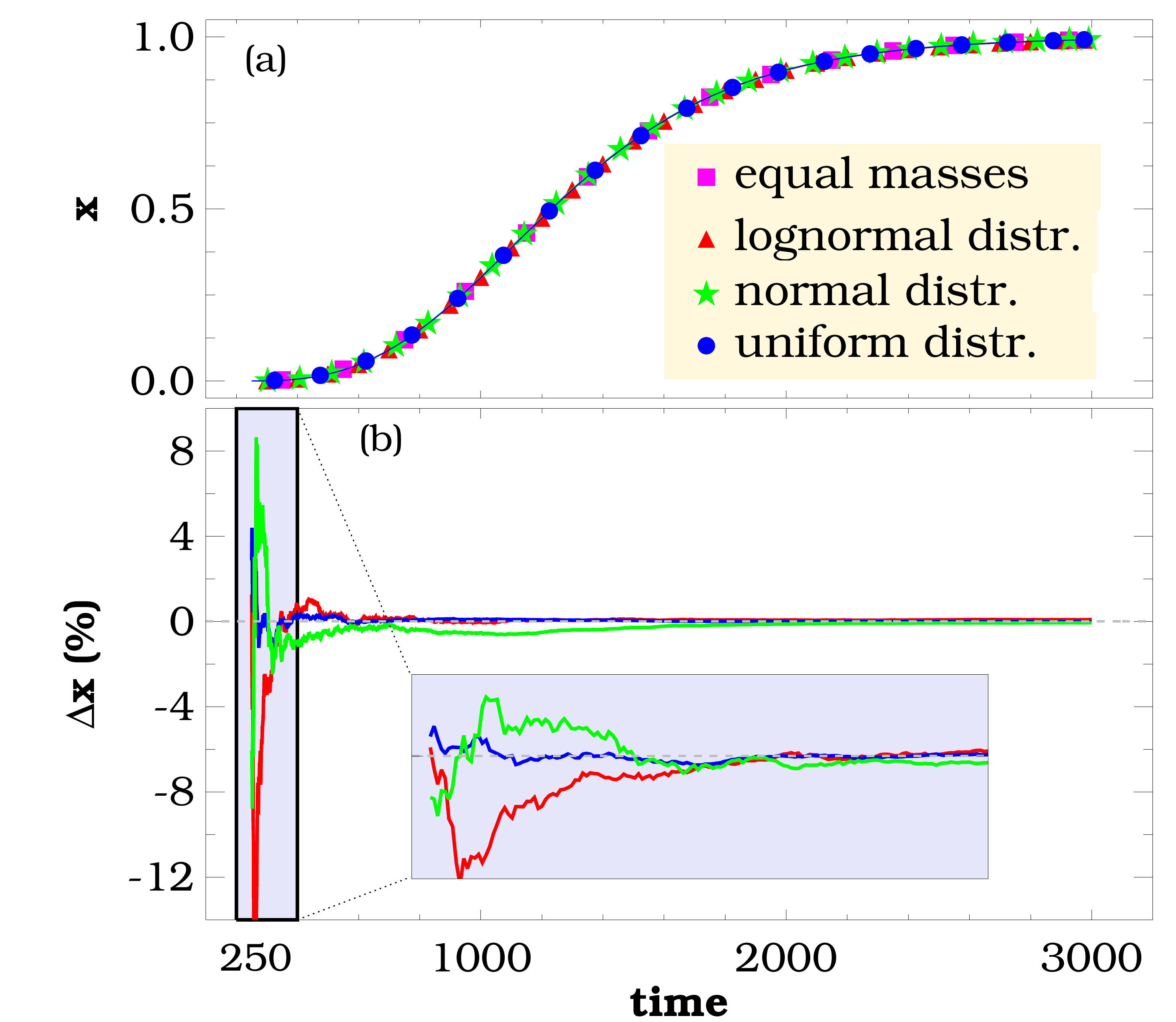}
\caption{\label{fig:logn}
(a) The leak mass vs. time for different mass distributions of particles each of them
with mean=1 and std=0.667.
Squares represent equal masses $m=1.$
(b) The differences between the leak's masses are significant only for the first ~200 iterations.
$K=2.7.$}
\end{figure}

In the previous two examples we considered particles with equal masses, $m=1.$
A more realistic scenario is when the particles in various physical problems
have different masses corresponding to a certain distribution.
The log-normal distribution is a good choice to describe the particle size (and/or mass).
We present a simulation for $\gamma=2/3$ with different kind of mass distributions, see Figure~\ref{fig:logn}.
The numerical results in panel (a) show what can also be derived directly from
Equations~(\ref{eq:N derivative}) and (\ref{eq:M derivative}):
\textit{the mass growth of the leak does not depend on the mass of the individual particles but only on the mean value of the distribution.}
Consequently, the leak's mass changes in time with the same rate
for both equal mass particles (pink squares) and log-normal distribution (red triangles),
and also for other distributions such as uniform and normal
(stars and circles in Figure~\ref{fig:logn}(a), respectively).
The statistical fluctuations in leak's mass, smaller than 15\%,
disappear after ~200 iterations, panel (b).

\section{Summary and Discussion\label{sec:sum_disc}}

The model Equations~(\ref{eq:N derivative}) and (\ref{eq:M derivative})
describe the escape dynamics in a leaky chaotic system when the size of the leak is growing in time and the expansion depends on the particles' mass.
Consequently, the escape probability is time-dependent.
The analytic solution to the problem provides a power-law behavior at the very early stage
($x\approx 0$) of the dynamical evolution. This phase depends on the exponent $\gamma$ in
Eq.~(\ref{eq:T_leak}).
However, for larger $t,$ when the feeding of the leak diminishes,
the survival decay turns to be exponential.
Between these two limits the escape rate is time dependent.

\begin{figure}[b]
\includegraphics[width=0.475\textwidth]{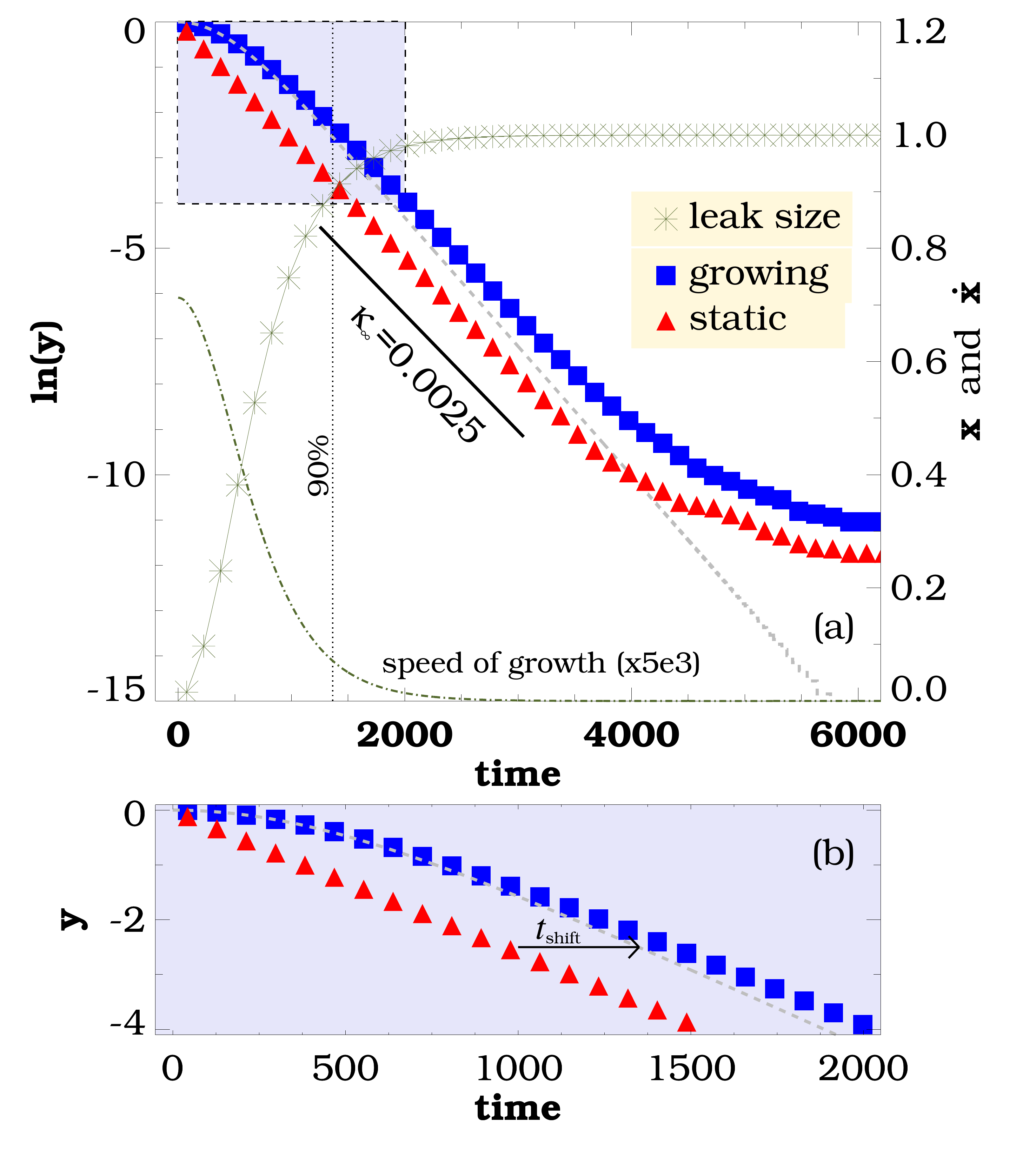}
\caption{\label{fig:holevelo}
%(a) The analytic solution Equation~(\ref{eq:sol}) is valid only for the beginning of
%simulations where the expansion of the leak is significant due to
%the large number of particles in the system.
%After this stage when the leak has almost the same size,
%the classical exponential decay appears and governs the escape.
(a) Number of non-escaped particles and leak size/growth vs. time for
$K=2.7,\;\;\gamma=1/2,\;\;m_{i}=1.$ (b) Magnification during growing
process.}
%The mass of the survivors is depicted together with the analytic solution.}
\end{figure}

The qualitative picture is the following.
After the leak reaches roughly the 90\% of its final measure, or more precisely,
beyond the point of inflection of $x(t)$, the speed of the growth slows down.
After this point the growth of the leak is so slow that it can be thought of
as a static leak, and the decay rate turns to be exponential,
see Figure~\ref{fig:holevelo}(a).
Numerical simulations verify that the escape rate $\kappa_{\infty}$
(short thick solid line) for a static leak (red triangles) of size 0.1 is the same
as in the case of a growing leak (blue squares) when it reaches 90\% of
its final size (also 0.1), panel (a).

In addition, this behavior is in a very good agreement with the analytical solution describing
the early stage escape dynamics.
{The effect is considerable for relatively short times only as long as enough number of
particles are in the system,
therefore, the presence of the well-known power-law decay of stickiness (tail of the distribution)
in mixed phase space is not affected by the size variation of the leak. However, the crossover time, when the nonhyperbolic part of the chaotic saddle starts to dominate, can be updated.}

{The crossover time $t_{\mathrm{cross}}$ in weakly chaotic regime is written as follows (Eq.~(89) in \cite{Alt2013})}
\begin{equation*}
t_{\mathrm{cross}}\sim 1/\kappa_{\infty} 
\end{equation*}
{with the assumption that the leak size is small. The growing leak model provides a simple generaliztion of this naive approximation in $\gamma\leq 1$ case}
\begin{equation}
t_{\mathrm{cross}}\sim \frac{1}{\kappa_{\infty}}\left[1+\frac{x^{1-\gamma}}{1-\gamma}+\frac{(1+\gamma)^{1+\gamma}}{\gamma^{\gamma}}\frac{\gamma}{1+\gamma}\right], 
\end{equation}
{where the second and the third terms in the bracket define the shift ($t_{\mathrm{shift}}$) the crossover experiences, see the shematic view in Fig.~\ref{fig:holevelo}b. The second term is the time of the growth until the leak mass is moderate, see the approximation of Eq.~(\ref{eq:ode for x}) when $x\ll1$, while the third term can be derived from the slope of the function $x(t)$ at point $x_{\mathrm{PoI}}$, Fig.~\ref{fig:analytical}. It can be easily obtained that $t_{\mathrm{shift}}\to 0$ when $\gamma\to 0$ and $x\ll1$.}

Equation~(\ref{eq:sol_gamma0}) properly describes also the limit case $m_{i}\to 0.$
Namely, if the mass of the particles tends to zero,
i.e. the growth of the leak is fairly slow,
one recovers the classical exponential decay for the surviving trajectories.
We note that the same effect can be seen when the initial mass of the leak $x_{0}$ is
set so large that even the massive particles ($m_{i}>0$) falling into it
do not have any effect on the leak's mass and, therefore,
it can be considered as a static leak.

{
Due to the leak expansion we can consider an instantaneous chaotic saddle in our model at every time step.
This object is reducing as the leak is growing and converges to that invariant set which corresponds to the final leak size.
This process results in a temporally changing chaotic saddle and a non-stationary exponent of the survival probability (escape rate $\kappa(t)$).
A similar phenomenon can be found in \cite{Mot2013} where the exponent is also time dependent (see Eq.~(1) in \cite{Mot2013}).
In contrast of the similarity, the temporarily changing chaotic saddle should not be confused with the transient chaotic saddle introduced in \cite{Mot2013}.}

In summary, we have presented an analytic description of the escape of
the trajectories through a continuously growing leak both in fully
hyperbolic and in mixed phase space.

We stress, however, that during the whole calculation we did not utilize explicitely the fact that $m$ is
the mass of the particle, though the basic motivation is related to the mass growth of
a planetary embryo.
Therefore, one can reformulate the model in a more general way.
{Let us write Eqs.~(\ref{eq:leakmass}), (\ref{eq:T_leak}), and (\ref{eq:escape_prob_M})
together as follows}
\begin{equation}
  \begin{split}
    M_i &= M_0 + (N_0-N_i) m = \mathcal{M} - N_i m,\\
    S_{\mathrm{leak},i}& = C_{S} M_i^\gamma,\\
    p_i &= C_p M_i^\gamma,
    \label{eq:escape_prob_M_general}
    \nonumber
%    \Phi &= \Phi_0 + (N_0-\bar N)\cdot \phi = \Xi - \bar N \cdot \phi,\\
%    S_{\mathrm{leak}} &= C_T \cdot \Phi^\gamma, \\
%    p &= C_p\cdot \Phi^\gamma, \label{eq:escape_prob_M_general}
  \end{split}
\end{equation}
{where now $m$ is a physical property of the particles, $M_n\text{ and } M_0$ are
the evolved and initial additive property of the leak, and $\mathcal{M} = M_0 + N_0 m.$}
Other quantities are the same as given in the introduction of the model in Section~\ref{sec:leak}.
This means that the analytical method presented in this paper might be suitable to predict the
characteristics of the escape dynamics in different kinds of systems where the leak size
depends on some specific physical property of the particles {(charge, spin, energy level, chemical composition, etc.)}.

We also would like to draw the attention to the limitation of present model.
In fact, the dynamics in the standard map does not depend on the size of the leak.
In other words, the leak affects only the escape rate but not the individual survival trajectories themselves.
This is not the case, for instance, in the restricted three body problem,
where the growing planetary mass governs the dynamics of the surviving particles and,
therefore, should also modify the escape dynamics. {Considering such an extension in the SM, a natural choice could be the introduction of a variable nonlinearity parameter $K(M)$ whose value could also depend on the leak mass/size.}
Studying this effect is postponed to future studies.

\appendix

\section{The exponent $\gamma$ in the planar RTBP \label{appendix:size_mass}}

%Our original motivation to the problem, the restricted three body problem
%implies a reasonable choice for the $S_{\mathrm{leak},n}\sim
%M_n^{\gamma}$ in Eq.~(\ref{eq:T_leak}).
In this section we show a short derivation for how the size of the leak in the RTBP
depends on the mass of the planetary embryo.
First, we can introduce a four-dimensional leak in the phase space of the RTBP.
Two dimensions out of four cover the physical extent of the planet
($0.5 r_{\mathrm{H}}$) in the configuration space, i.e. the small gray circle at the position (-1,0) in Figure \ref{fig:rtbp}.
The remaining two components whose absolute value is the escape velocity at half of
$r_{\mathrm{H}}$ describe the size of the leak in the velocity space.
In fact, the Hill radius and the escape velocity, as described above, can be {written as a function} of the planet's mass.
Hence, the size of the 4 dimensional leak in phase space depends only on the mass of the
planet $(m_p).$
%The derivation of the exact formula of $S_{\mathrm{leak}}(M)$ in RTBP is postponed to the Appendix \ref{appendix:size_mass}.

The Hill radius $r_{\mathrm{H}}$ is defined
\begin{equation}
r_{\mathrm{H}}=a\left(\frac{\mu}{3}\right)^{1/3} \label{eq:rH}
\end{equation}
where $\mu=m_{\mathrm{p}}/M_{\mathrm{s}}$ is the planet-to-star mass ratio and $a$ is the planet's semi-major axis.
In addition, a particle must have a smaller velocity than the escape velocity in
order to be trapped in a pre-defined region, e.g. in one half of the Hill radius.
The escape velocity from $0.5 r_{\mathrm{H}}$ reads
\begin{equation}
v_{\mathrm{esc}}=\sqrt{\frac{4Gm_{\mathrm{p}}}{r_{\mathrm{H}}}} \label{eq:vEsc}
\end{equation}
where $G$ denotes the gravitational constant and $m_{\mathrm{p}}$ is
the planetary embryo's mass.

Thus, the size of the leak ($S_{\mathrm{leak}}$) in the phase space of the RTBP is obtained
as the product of the spatial ($A_{\mathrm{r}}=\pi
r_{\mathrm{H}}^{2}$) and velocity extensions ($A_{\mathrm{v}}=\pi v_{\mathrm{esc}}^{2}).$
That is, we have a leak size with $\gamma=4/3$
\begin{equation}
S_{\mathrm{leak}}=A_{\mathrm{r}} A_{\mathrm{v}}\sim r_{\mathrm{H}}^{2} v_{\mathrm{esc}}^{2}\propto m_{\mathrm{p}}^{4/3}.
\label{eq:hole2D}
\end{equation}

\section{Solution of Eq.~(\ref{eq:ode for x}) \label{appendix:solution_of_eq}}

Let us recall Eq.~(\ref{eq:ode for x})
\begin{eqnarray}
\frac{\mathrm{d}x}{\mathrm{d}t} = \kappa_{\infty} \cdot x^\gamma \cdot \left( 1 - x \right),
\label{appeq:ode for x}
\end{eqnarray}
After arrangement and integration we have
\begin{eqnarray}
\frac{1}{\kappa_{\infty}}\int\frac{1}{x^\gamma}\cdot \frac{1}{1-x} \; \mathrm{d}x =
\int 1\; \mathrm{d}t. \label{eq:integral}
\end{eqnarray}
In the special case of $\gamma=1$
\begin{eqnarray}
\int\frac{1}{x}\cdot \frac{1}{1-x}\; \mathrm{d}x = \ln \frac{x}{1-x},
\label{eq:sol gamma 1}
\end{eqnarray}
and
\begin{eqnarray}
x(t)=\frac{1}{ 1 + e^{-\kappa_{\infty} (t+\tau)}},
\label{eq: sol gamma 1 exp}
\end{eqnarray}
where
\begin{eqnarray}
\tau=\ln \frac{x_0}{1-x_0}
\label{eq: sol gamma 1 exp tau}
\end{eqnarray}
is the constant of integration.
In the case of $\gamma \ne 1$, first, we consider the fact that
\begin{eqnarray}
\frac{1}{1-x} =
\sum_{i=0}^{\infty}x^i.
\end{eqnarray}
Now, the integral on the LHS of Eq.~(\ref{eq:integral}) can be written as
\begin{eqnarray}
\int\frac{1}{x^\gamma}\cdot \frac{1}{1-x}\; \mathrm{d}x = \int
\sum_{i=0}^{\infty} x^{i-\gamma} \; \mathrm{d}x= \nonumber
\\
x^{1-\gamma}\sum_{i=0}^{\infty} \frac{x^i}{i-\gamma+1} =
\frac{x^{1-\gamma}}{1-\gamma}\sum_{i=0}^{\infty}
\frac{(1-\gamma)^{(i)} \cdot 1^{(i)}}{(2-\gamma)^{(i)}}
\frac{x^i}{i!} = \nonumber
\\
\frac{x^{1-\gamma}}{1-\gamma} \cdot
{}_2F_1\left(1-\gamma,1;2-\gamma;x\right) \nonumber
\end{eqnarray}
where $q^{(i)}$ is the rising Pochhammer symbol
\begin{eqnarray*}
q^{(i)} = q (q+1)\dots (q+i-1)
\end{eqnarray*}
and
%where %${}_2F_1$ denotes the Gaussian hypergeometric function.
\begin{eqnarray*}
{}_2F_1(a,b;c;x)= \sum_{i=0}^{\infty}
\frac{a^{(i)}\;b^{(i)}}{c^{(i)}}\; \frac{x^i}{i!}
\end{eqnarray*}
is the Gaussian hypergeometric function \cite{AS1970,RG2014}.
Taking $1/\kappa_{\infty}$ on the LHS and performing the integration on the RHS,
the solution as given by Eq.~(\ref{eq:sol}) is obtained.
%\begin{eqnarray}
%\frac{x^{1-\gamma}}{\kappa_{\infty} \cdot (1-\gamma)}\;
%{}_2F_1\left(1-\gamma,1;2-\gamma;x\right) = t+\tau \label{eq:sol}
%\end{eqnarray}
%where the constant of integration $\tau$ follows from the initial value $x_0$ as
%\begin{eqnarray}
%\tau = \frac{x_0^{1-\gamma}}{\kappa_{\infty} \cdot (1-\gamma)}\;
%{}_2F_1\left(1-\gamma,1;2-\gamma;x_0\right).
%\label{eq:int_const}
%\end{eqnarray}

For certain rational $\gamma$ values the implicit solutions of Eq.~(\ref{eq:ode for x})
(corresponding to the integral on the left-hand side of
(\ref{eq:integral})) are summarized in Table~\ref{table:gamma}.
%  can be expressed by elementary functions.
%shows these functions for different values of $\gamma.$

\begin{table}[b]
\caption{\label{table:gamma}
The integral on the left-hand side of (\ref{eq:integral}) expressed by elementary
functions.
}
\begin{ruledtabular}
\begin{tabular}{cccc}
$\gamma$ & $\int\frac{1}{x^\gamma}\cdot \frac{1}{1-x}\; \mathrm{d}x$ \\
\colrule
0   & $ - \ln\left( 1 - x \right)$ \\
  1/2 & $2 \tanh^{-1}\left(\sqrt{x}\right)$ \\
  3/4 & $
  \ln\left( \frac{1+\sqrt[4]{x}}{1-\sqrt[4]{x}} \right)
  +
  2 \tanh^{-1}\left(\sqrt[4]{x}\right)
  $ \\
  1   & $\ln\left( \frac{x}{1-x} \right)$ \\
  4/3 & $
  \ln\left(\frac{ \sqrt{1+\sqrt[3]{x}+\sqrt[3]{x^2}}}{1-\sqrt[3]{x}}\right)
  - \sqrt{3} \tanh^{-1}\left(\frac{1+2 \sqrt[3]{x}}{\sqrt{3}}\right)
  -\frac{3}{\sqrt[3]{x}}
  $ \\
  3/2 & $ 2 \tanh^{-1}\left(\sqrt{x}\right) - \frac{2}{\sqrt{x}}$ \\
  2   & $\ln\left( \frac{x}{1-x} \right)-\frac{1}{x}$ \\
\end{tabular}
\end{ruledtabular}
\end{table}

Interestingly, in addition to $\gamma=1,$ the solutions for $\gamma=0$ and $\gamma=1/2$ can also be given in explicit forms as follows
\begin{eqnarray}
x(t) = 1 - (1-x_0) \cdot %- \left( 1 - x_0 \right)
e^{- \kappa_{\infty} t}
\label{eq:sol_gamma0}
\end{eqnarray}
and
\begin{eqnarray}
x(t) = \tanh^2 \left[\frac{ \kappa_{\infty}(t+\tau) }{2}\right],
\label{eq:sol_gamma1_2}
\end{eqnarray}
respectively.
Equation (\ref{eq:sol_gamma0}) provides the classical exponential decay for $\gamma=0,$ when the leak is stationary.

\section{Error analysis \label{appendix:error}}

%The Eq. (\ref{eq:ode for x}) is a continuous approximation of the discrete
%sequence given by the recursive formula

%During the simulation, the discrete-time chain of the averaged mass ratios is
%a sequence $(x_i)^\infty_{i=0}$ governed by the recursive formula
During the simulation, the sequence of the averaged mass ratios $(x_i)^\infty_{i=0}$ is
governed by the recursive formula (\ref{eq_discrete}).
%\begin{equation}
%%\Delta x_i = \kappa_\infty \cdot x_i^\gamma \cdot \left( 1 - x_i \right)
%x_{i+1}=x_i + \kappa_\infty \cdot x_i^\gamma \cdot \left( 1 - x_i \right).
%\label{eq_discrete}
%\end{equation}
The differential equation (\ref{eq:ode for x}) and its implicit solution (\ref{eq:sol}) give only a continuous approximate solution of the original discrete problem.
The question arises naturally, how good the approximation (\ref{eq:sol}) is?

\begin{figure}
\includegraphics[width=0.475\textwidth, angle=0]{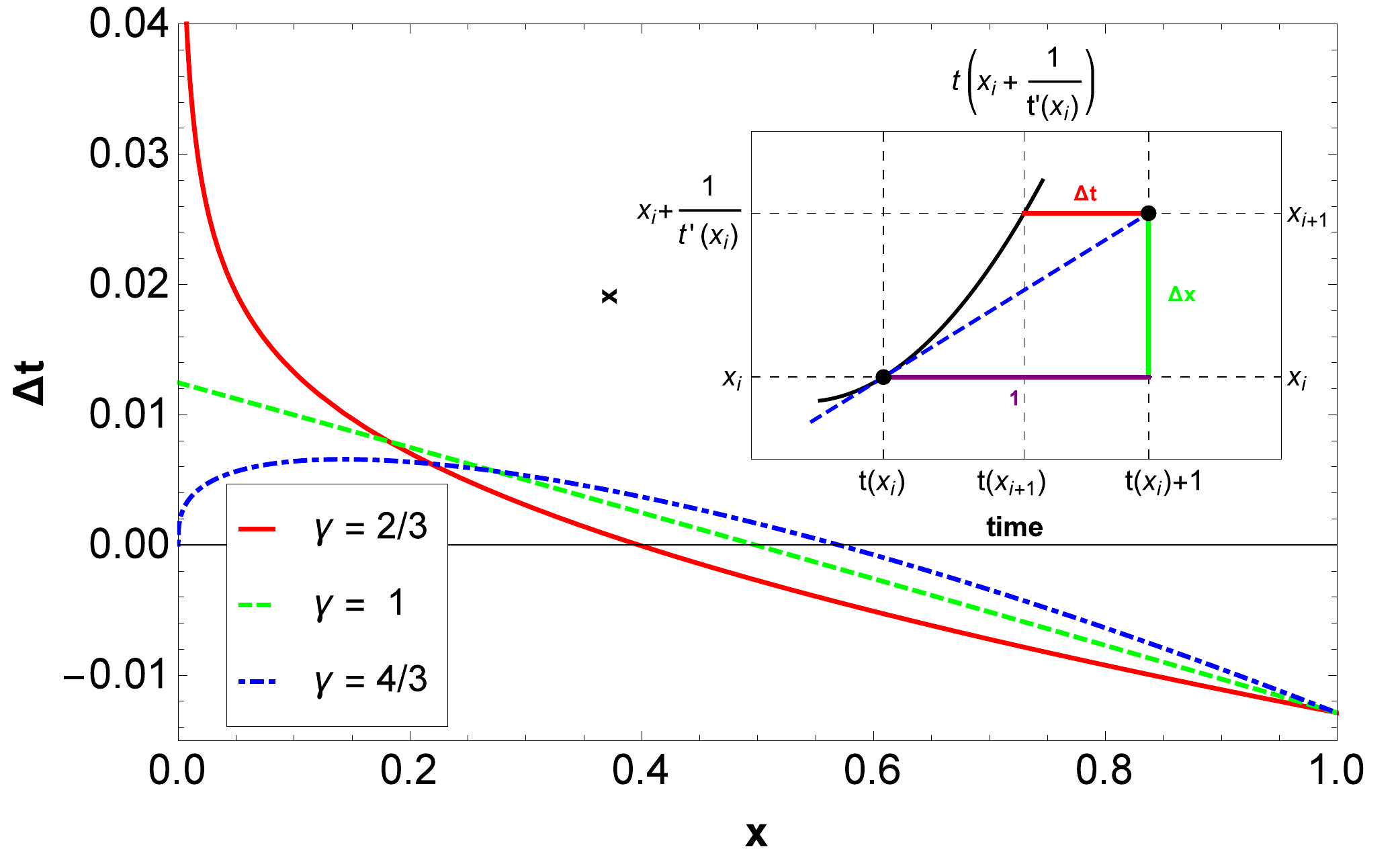}
\caption{(color online)
The relative error of the approximation Eq.~(\ref{eq_deltat}) after a single iteration as a function of the mass ratio $x$.
%The inset shows the function represented by Eq.~(\ref{eq_deltat}).
%The relative error of the approximation caused by one iteration via massratio $x$.
%The inset helps to understand the formula (\ref{eq_deltat}).
The inset helps to understand the formula of the relative error.
The black dots represent the two successive terms.
The black solid line and the (blue - online) dashed line display the continuous solution $t(x)$ %($x(t)$)
and its tangent curve, respectively.
The parameters are the same as in Fig. \ref{fig:quantiles_all}.
\label{fig_differences}
}
\end{figure}

Let us consider two successive terms of the original sequence $x_i$ and $x_{i+1}$
(see the inset of Fig. \ref{fig_differences}).
According to the approximation $t(x)$, %which is known,
the time interval between the two states is $t(x_{i+1})-t(x_i)$ instead of 1.
The difference $\Delta t(x_i) = 1 - \left[ t(x_{i+1}) - t(x_i) \right]$ is the (relative) error
%and also the relative error
of the approximation caused by one iteration.
%As $x_{i+1} = x_i + \frac{1}{t'(x_i)}$
As $x_{i+1} = x_i + 1/t'(x_i)$,
%(do not forget that $\Delta x = x_{i+1} - x_i = 1/t'(x_i)$),
%(see the inset of Fig. \ref{fig_differences}),
function $\Delta t$ can be expressed as
\begin{equation}
\Delta t\left(x\right) = 1 - \left[ t\left( x + \frac{1}{t'(x)}\right) - t(x) \right].
\label{eq_deltat}
\end{equation}
Figure \ref{fig_differences} shows the functions $\Delta t(x)$ for different $\gamma$s.
%The error function changes its sign at the value of $x_{PoI}$.
In the cases of $\gamma = 4/3$ and $\gamma = 1$ the relative errors remain unter $1.2\%$ (in general under $\kappa_{\infty}/2$).

Unfortunately, in the third case ($\gamma=2/3$),
$\lim_{x \to 0} t(x) = 1$ (100\% relative error).
In small $x$ approximation, %(\ref{eq_discrete})
more precisely if $x_i \ll \kappa_\infty^{\frac{1}{1-\gamma}}$
%(in the simulation $\kappa_\infty^{\frac{1}{1-\gamma}} \approx 1.63 \cdot 10^{-5}$)
,
the recursive formula (\ref{eq_discrete}) can be approximated by
$x_{i+1} \approx \kappa_\infty \cdot x_i^\gamma$.
This recursive sequence can be written in explicit form as
\begin{equation}
x_i=\kappa_\infty^{\frac{1}{1-\gamma}} \cdot
\left(x_0 \cdot \kappa_\infty^{\frac{1}{\gamma-1}}\right)^{\gamma^i}.
\end{equation}
This sequence is increasing really fast from any astronomically small value $x_0$ to
$x_i \approx \kappa_\infty^{\frac{1}{1-\gamma}}$.
%Ez a megoldás egy rendkívül gyors, rövid idejű növekedést jelent bármilyen
%csillagászatian kicsi tömegarányról $x_i \approx \kappa_\infty^{\frac{1}{1-\gamma}}$-ra.
If $x_0 = 10^{-a} \cdot \kappa^{\frac{1}{1-\gamma}}_\infty$
and $x_i = 10^{-b} \cdot \kappa^{\frac{1}{1-\gamma}}_\infty$
then the time period of the growing is
\begin{equation}
i=
%\frac{\log\left[\log\left( x_i \cdot \kappa^{-\frac{1}{1-\gamma}}_\infty\right)\right]
%-\log\left[\log\left( x_0 \cdot \kappa^{-\frac{1}{1-\gamma}}_\infty\right)\right]}
%{\log(\gamma)}=
\frac{\log a-\log b}{\log(\gamma)}.
\label{eq_loglog}
\end{equation}
For example, in the case of
$x_0 = 10^{-100} \cdot \kappa^{\frac{1}{1-\gamma}}_\infty$ ($a=100$)
and
$x_i = 0.98 \cdot \kappa^{\frac{1}{1-\gamma}}_\infty$ ($b\approx0.01$), $i \approx 23$.
The continuous approximation does not describe this fast growing process.

Although the relative error decreases under $3.7\%$
at $x\approx0.01$ (see Fig. \ref{fig_differences}),
according to our numerical results the global error is acceptable
if the initial mass ratio $x\gtrapprox \kappa_\infty^{\frac{1}{1-\gamma}}$.
%The reason of this is that
For example, in the corresponding case of Fig. \ref{fig:quantiles_all}
in spite of the initial mass ratio ($x_0=10^{-6}$) is slightly smaller than
$\kappa_\infty^{\frac{1}{1-\gamma}}$,
the global error remains under $4\%$.

\section{Distributions \label{appendix:distr}}

Let $\xi_i$ be discrete random variables associated with the number of particles after
the $i$th iteration.
Here we derive the probability mass function $P(\xi_{i+1}=k),\;k=0\dots N_0$ by assuming
that it is known from the earlier iterations $P(\xi_i=j),\;j=0\dots N_0.$
%We give the solution to the problem by answering the question
%how the $(i+1)$th probability density function can be derived from the $i$th one.
%In other words, if the probabilities $P(\xi_i=j),\;j=0\dots N_0$ are known then what are
%$P(\xi_{i+1}=k),\;k=0\dots N_0?$

The number of escaping particles during one iteration follows binomial distribution.
Let us suppose that there are $j$ particles in the system ($\xi_i=j$) and after
one iteration the number of particles is $k\le j$ ($\xi_{i+1}=k$),
then the number of escaping particles is $j-k.$
Using the formula of the binomial distribution,
we can write the following conditional probability
\begin{eqnarray}
P(\xi_{i+1}=k \;|\; \xi_i=j) = \binom {j} {j-k}\; p_j^{j-k} (1-p_j)^{k},\;\;\;
\label{eq:conditional_probabilities}
\end{eqnarray}
where $p_j$ is the escape probability which corresponds to the particle number
$N = j,$ namely
\begin{eqnarray}
p_j = C_p\cdot [M_0 + (N_0-j)\cdot m]^\gamma.
\label{eq:recursive}
\end{eqnarray}
According to the law of total probability, we can write
\begin{eqnarray}
P(\xi_{i+1} = k)=\sum_{j=k}^{N_0} P(\xi_{i+1}=k \;|\; \xi_i=j)\cdot
P(\xi_i = j),\;\;\;\;\;\;
\label{eq:law_of_total probability}
\end{eqnarray}
thus we get a recursive formula for $P(\xi_{i+1} = k).$
If the initial number of particles is set to be $N_0$ then the initial distribution reads
\begin{eqnarray}
P( \xi_0 = k )= \left\{
 \begin{array}{ll}
  1 &\;\;\; \textrm{if $k  =  N_0$,}\\
  0 &\;\;\; \textrm{if $k \ne N_0$,}
 \end{array}
\right.
\label{eq:initial distributione}
\end{eqnarray}
and any $P(\xi_{i+1} = k)$ probability can be calculated recursively
by using (\ref{eq:conditional_probabilities})--(\ref{eq:initial distributione}).

\begin{figure}
\includegraphics[width=0.95\linewidth, angle=0]{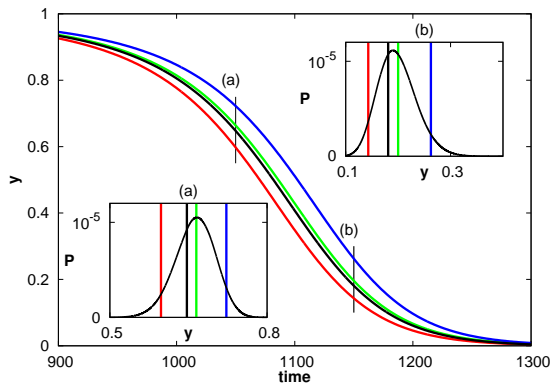}
\caption{
The 1st (red), the 5th (green), and the 9th (blue) deciles of the series of distributions
in the case of $\gamma=4/3.$
%Parameters: $N_0=10^6$, $m=1$, $M_0=1000$, and $C_p=(2 \pi)^{-2} \cdot N_0^{-\gamma}.$
To make the distinctions of the three curves easier, we cut off their first parts,
$t<$900.
The black curve shows the particle number $y(t).$
The insets (a) and (b) show two distributions corresponding to the iterations indicated
by the two vertical black lines.
\label{fig:quantiles_gamma43}}
\end{figure}

In order to check whether the analytic model is valid, several calculations of distribution series were carried out.
Figure \ref{fig:quantiles_gamma43} shows the 1st, 5th, and 9th deciles (10-quantiles) of
the series of distributions for $\gamma=4/3.$
This calculation is suitable to test the accuracy of the particle number ratio %number
$y(t) = 1-x(t)$
calculated in the section \ref{sec:averaged_behaviour}.
The analytic solution is also plotted (black curve) together with the statistical results.
One can see that the function $y(t)$ is close the decile curves which means that
the analytic solution is suitable to approximate the discrete process.

\begin{figure}
\includegraphics[width=0.475\textwidth, angle=0]{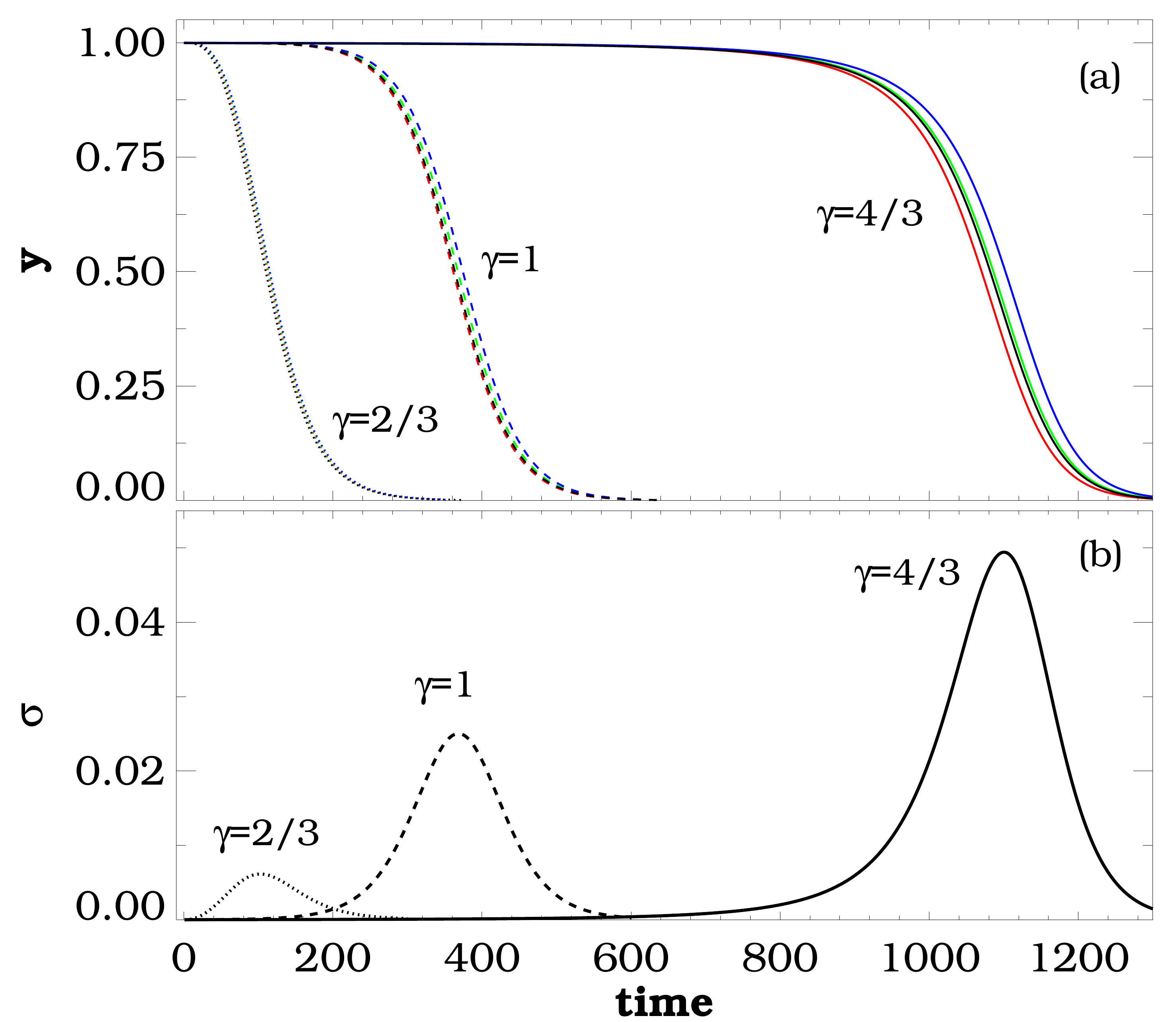}
\caption{
(a) The 1st (red), the 5th (green), and the 9th (blue) deciles of the three calculated series of distributions
%($\gamma=2/3$ - dotted, $\gamma=1$ - dashed, $\gamma=4/3$ - solid).
%The initial masses of the leaks are $x_0=1$ ($\gamma=2/3$), $x_0=100$ ($\gamma=1$),
%and $x_0=1000$ ($\gamma=4/3$).
%The other parameters are $N_0=10^6$, $m_{i}=1$, and
%$C_p=(2 \pi)^{-2} \cdot N_0^{-\gamma}.$
The black curves show the particle number ratio ($(1-x)$) calculated
in the section \ref{sec:averaged_behaviour}.
(b) Standard deviations of the distributions for different $\gamma$s.
%$\gamma=2/3$ - dotted,
%$\gamma=1$ - dashed, $\gamma=4/3$ - solid.
\label{fig:quantiles_all}}
\end{figure}

We also verified these result by analyzing distributions for different $\gamma$s.
Figure~\ref{fig:quantiles_all} illustrates the results
%for $\gamma=2/3$, $\gamma=1$, and $\gamma=4/3$.
for $\gamma=2/3,\;1$, and $4/3$.
%The initial masses of the leaks were $x_0=10^{-6},\;10^{-4}$, and $10^{-3}$, respectively.
The other parameters were $N_0=10^6$, $m=1$, and
$C_p=(2 \pi)^{-2} \cdot N_0^{-\gamma}$ in all three cases
(the initial escape probabilities were the same, $p_0=10^{-4} \cdot (2\pi)^{-2}$
and $\kappa_{\infty} \approx 1/(2\pi)^{2}=2.53\cdot10^{-2}$).
The distributions were calculated until their averages decreased
under 0.1 percent of the initial particle number ($E(\xi_i)<10^{-3}\cdot N_0$).
Figure \ref{fig:quantiles_all}(b) shows the standard deviations in all three cases.
In general, the standard deviations are not negligible but remain relatively small.

\begin{acknowledgments}
We are indebted to G. Kov\'acs and T. T\'el for useful discussions. The authors also thank the
anonymous referees their valuable comments and suggestions that
helped to improve the text significantly.
This work was partially supported by the OTKA Grant No. NK100296,
K119993, and PD121223.
TK also thanks for the support for the Fulbright Commition and the Hungary
Initiatives Foundation.
\end{acknowledgments}

%\bibliography{literature}% Produces the bibliography via BibTeX.
\newcommand{\apjl}{Astrophysical Journal Letters}\newcommand{\mnras}{Monthly
  Notices of the Royal Astronomical Society}
\end{document}